\documentclass[preprint]{ptephy_v1}

\preprintnumber{XXXX-XXXX} 

\usepackage{amsmath}
\usepackage{amsthm}
\usepackage{hyperref}
\usepackage{url}

\usepackage[T1]{fontenc} 
\usepackage[utf8]{inputenc}
\usepackage{bm}

\begin{document}

\title{Competition of the shell closure and deformations across the doubly magic \textsuperscript{78}Ni}


\author{Ryo Taniuchi}\affil{School of Physics, Engineering and Technology, University of York, York, YO10~5DD, United Kingdom \email{ryo.taniuchi@york.ac.uk}}


\begin{abstract}%
The properties of the neutron-rich isotope \textsuperscript{78}Ni, long postulated to be doubly magic, have been extensively explored through recent experimental and theoretical studies. Confirmations of robust shell closures at $Z=28$ and $N=50$ as well as hints of competing deformations in neighboring isotopes have been obtained. Innovations of a thick liquid hydrogen target system with vertex reconstructions and the in-beam $\gamma$-ray spectroscopy technique have facilitated detailed investigations into the nuclear structure of these extreme systems. Proton knockout reactions conducted at relativistic energies have provided the first experimental evidence of shape coexistence at the cornerstone nucleus \textsuperscript{78}Ni and its vicinity. As the nuclear structure around \textsuperscript{78}Ni influences the description of very neutron-rich systems and r-process nucleosynthesis, these findings underscore the importance of further investigations. This review encapsulates the recent results concerning the nuclear structure at the vicinity of \textsuperscript{78}Ni on both experimental and theoretical aspects. It outlines prospective research directions that could further illuminate this complex and intriguing area of the nuclear chart.


\end{abstract}

\subjectindex{xxxx, xxx}

\maketitle

\section{Introduction}

Atomic nuclei consist of protons and neutrons and are successfully described in a picture of independent particles in a mean-field potential with additional spin-orbit interactions.
In recent decades, many experimental results have shown the model is not any more robust towards neutron-rich isotopes; in systems with large neutron-to-proton imbalance, canonical magic numbers disappear and new ones emerge.
Such phenomenon is known as shell evolution; the effective single-particle orbitals of protons and neutrons vary and re-order.

\subsection{Competitions between shell-closures and shape coexistence}

The area around \textsuperscript{78}Ni has been intensively investigated as it is expected to be a region of competition between the regimes of spherical mean field and nuclear correlations~\cite{Nowacki2021}.
While \textsuperscript{78}Ni had been believed to be a doubly magic nucleus with a rigid spherical shape, evidence shows that the $N=40$ Island of Inversion (IoI) extends towards $N=50$~\cite{Nowacki2021, Santamaria2015}.
As \textsuperscript{78}Ni serves as a foothold for describing further neutron-rich nuclei involved in r-process nucleosynthesis, a precise understanding of this keystone nucleus is essential.
Since its first observation in a bound state at the fragment separator FRS at GSI~\cite{Engelmann1995}, it took a decade to achieve the first measurement of the $\beta$-decay lifetimes of \textsuperscript{78}Ni and its vicinity~\cite{Hosmer2005}. Nevertheless, detailed investigations of this isotope have remained challenging until recently.

The proton single-particle states above the proton $Z=28$ shell-gap have been examined using odd-mass copper isotopes.
An earlier work on $\beta$-$\gamma$ spectroscopy of the \textsuperscript{68-74}Cu isotopes manifested the gradual reduction of the energy of the $I^\pi=f_{5/2}$ states towards $N=50$~\cite{Franchoo1998}.
The ordering of the $f_{5/2}$ and $p_{3/2}$ states was found to be swapped at \textsuperscript{75}Cu ($N=46$)~\cite{Flanagan2009}.
As the reconfiguration would lead to the weakening of $Z=28$ gap towards $N=50$, further detailed studies have been conducted intensively such as $\gamma$-ray spectroscopy experiments of the neutron-rich zinc~\cite{VanDeWalle2007,Shiga2016} and copper~\cite{Sahin2017,Vajta2018} isotopes.

The neutron numbers $N=40$ and $N=50$ are recognized as regions of shell evolution in the neutron-rich nickel region.
While clear evidence of the extension of IoI was found in the cobalt and iron isotopes extending the $N=40$ IoI towards the $N=50$ gap~\cite{Santamaria2015, CortesSua2024}, the nickel isotopes seemingly sustain the shell closure.
\textsuperscript{68}Ni is particularly noteworthy for exhibiting features of shape coexistence, which keeps the shell closure in the ground state as highlighted in previous researches~\cite{Suchyta2014, Tsunoda2015}.

The persistence of the shell-closures in proximity to \textsuperscript{78}Ni has been questioned by several experimental and theoretical works.
Collinear laser spectroscopies on the \textsuperscript{78}Zn isotope revealed that the charge radius of the long-lived isomer ($I^\pi = 1/2^+$) state is larger than that of the ground ($I^\pi = 9/2^+$) state~\cite{Yang2016}.
In \textsuperscript{80}Zn, a direct feeding to the ground state from a 2627(39)-keV state was observed~\cite{Shiga2016}. Although no spin or parity was assigned in the study, it can be considered as a decay from a possible ($2^+_2$) state if the shape coexistence prevails in this region.
The presence of the long-lived isomer in $^{79}$Zn at 943~keV was further confirmed by a $\beta$-delayed $\gamma$-ray spectroscopy~\cite{Marie-Coralie-Thesis} and precise mass measurements~\cite{Nies2023}.
These results demonstrate the manifestation of coexisting shapes in the area of the doubly-magic nuclei and underscore the importance of further experimental input on \textsuperscript{78}Ni and \textsuperscript{79}Cu.

\subsection{Experimental studies towards \textsuperscript{78}Ni}
\begin{figure*}
    \centering
    \includegraphics[width=\linewidth]{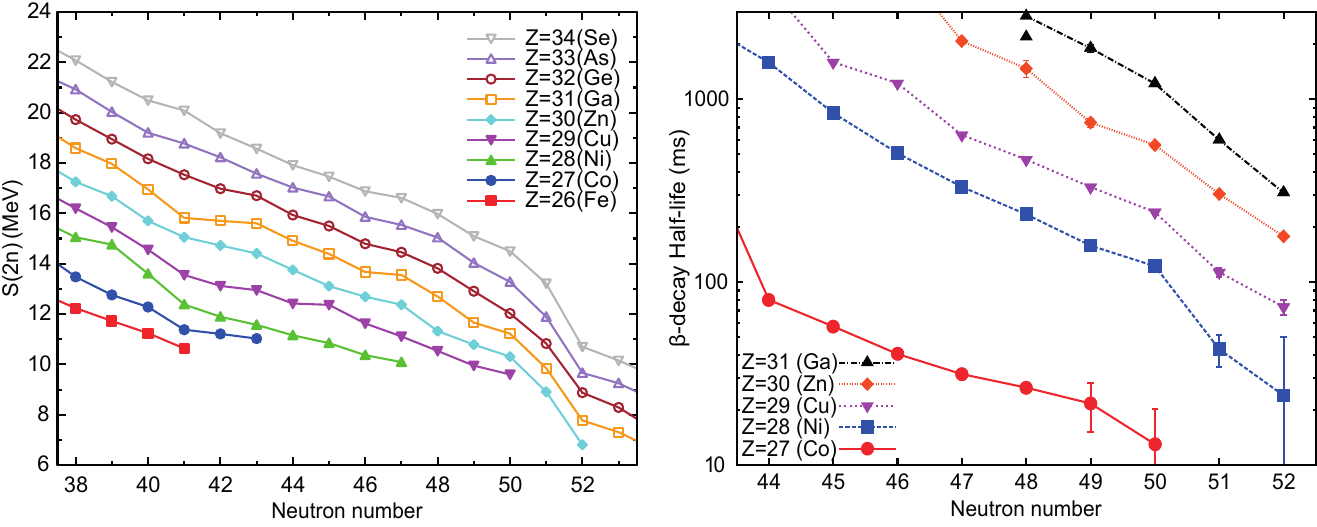}
    \caption{%
    \textbf{Left:} Two-neutron separation energies $S_\textrm{2n}$ towards \textsuperscript{78}Ni. Values are taken from the atomic mass evaluation, AME2020~\cite{Wang2021} and recently reported values with penning trap experiments~\cite{Giraud2022}.
    \textbf{Right:} Experimental data for the half-lives around \textsuperscript{78}Ni. The gap between cobalt and nickel isotopic chains and the change in the kinks across $N=50$ indicate the shell-closures of proton- and neutron-shells~\cite{Xu2014, NNDC}.%
    }
    \label{fig:systematics}
\end{figure*}

There are several approaches to evaluating the shell gaps.
Precise mass measurements, which provide the binding energy of the valence nucleons, are considered as the most direct observations.
The systematic trend of the two neutron/proton separation energies $S_{\textrm{2n/2p}}$\footnote{%
It is defined as 
$S_{\textrm{2n}}(N,Z) = \textrm{BE}(N,Z) - \textrm{BE}(N-2,Z)$, where $\textrm{BE}(N,Z)$ represents the binding energy. To identify shell gaps, the difference between the two-neighboring $S_{\textrm{2n}}$, $\Delta_{2n}(Z,N) = S_{\textrm{2n}}(N,Z) - S_{\textrm{2n}}(N + 2,Z)$ is also often used.
} is a direct measure of the gaps.
The left panel of Figure~\ref{fig:systematics} shows the $S_{\textrm{2n}}$ values deduced from the measured masses around \textsuperscript{78}Ni~\cite{Wang2021,Giraud2022}.
Recently, penning trap and MR-TOF mass spectroscopies at IGISOL-Jyv\"askyl\"a and CERN-ISOLDE~\cite{Welker2017, Giraud2022}, reached the measurements to \textsuperscript{79}Cu and \textsuperscript{75}Ni, while the masses beyond \textsuperscript{75}Ni is still unreached.

To assess the shell gaps, measuring $\beta$-decay lifetimes is an alternative approach.
The transition probability is correlated with the $\beta$-decay $Q_\beta$ values.
The shell-closure at \textsuperscript{78}Ni is indicated by the sudden reduction of the lifetimes beyond $N=50$ in the systematic trend~\cite{Xu2014, NNDC}, as shown in the right panel of Figure~\ref{fig:systematics}.
However, the lifetime also depends on the matrix elements of $\beta$-decay transitions, meaning the result cannot be considered conclusive.

\begin{figure}
    \centering
    \includegraphics[width=\textwidth]{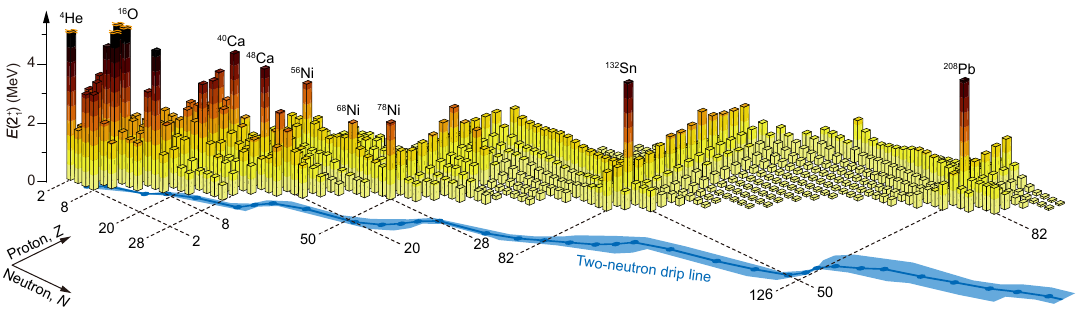}
    \caption{The $E(2^+_1)$ systematics of even-even isotopes. Canonical magic numbers are indicated by dashed lines. The predicted two-neutron drip line, that is the boundary of the existence of isotopes, and its uncertainties are also illustrated with the blue line and shaded area~\cite{Erler2012}. The relatively large excitation energies can be confirmed in the closed-shell nuclei. This figure is the updated version of ref.~\cite{Taniuchi2019} with values from NNDC~\cite{NNDC} and recent measurements~\cite{Taniuchi2019, Chen2023, Cortes2020,Liu2019}.
    }
    \label{fig:E2 Systematics}
\end{figure}

The systematic trend of the $2^+$ excitation energies of even-even isotopes is a good indicator of the shell closure of nuclei and is often served as the first observation.
Figure~\ref{fig:E2 Systematics} shows the systematic chart of the experimentally known $E(2^+_1)$ values. The canonical magic numbers are valid throughout the nuclear chart in the heavier regions, especially at \textsuperscript{132}Sn and \textsuperscript{208}Pb, while reconfiguration of magic numbers can be seen in the lighter region, such as the disappearance of $N=20$ and $N=28$ shells in the IoI region and the appearance of new magic numbers $N=32$ and $N=34$ for \textsuperscript{52,54}Ca.

At the RIBF, intensive spectroscopic studies far off the stability line have been achieved with the combination of the NaI(Tl) scintillator-based high-efficiency $\gamma$-ray spectrometer, DALI2~\cite{Takeuchi2014}, and the high luminosity liquid hydrogen target system with the capability of vertex reconstructions based on a coaxial time-projection chamber (TPC), MINOS~\cite{Obertelli2014}.
The experiment took advantage of the in-beam $\gamma$-ray spectroscopy technique to populate and measure the de-excitation $\gamma$ rays from excited states of short-lived isotopes using the inverse kinematics.
In this review, experimental results from the first experimental campaign employing MINOS with DALI2, which took place in 2014 aiming at the region of \textsuperscript{78}Ni~\cite{Doornenbal2015a}, and the relevant theoretical developments are discussed. Noteworthy, except for a few isotopes around the $N=40$ region, all the experimental outcomes addressed in this review were obtained by a single secondary-beam setting measured for about 6.5 days beamtime with a high-intensity \textsuperscript{238}U beam (12~pnA at the time of the experiment) at the RIBF facility.

\section{The first spectroscopy experiments in the vicinity of \textsuperscript{78}Ni}
The nucleus $^{78}$Ni holds a particular interest in nuclear physics as a potential doubly magic isotope, situated at far neutron-rich from the last stable isotope $^{64}$Ni.
The persistence of magic numbers in nuclei, especially near neutron-rich extremes, is challenging for the theoretical descriptions of nuclear structures and shell evolution.
Previous theoretical studies, utilizing both shell model calculations and first-principles approaches, have predicted $^{78}$Ni to exhibit characteristics of a doubly magic nucleus, while several experimental observations cast questions on this statement.
Experimental investigations in the vicinity of \textsuperscript{78}Ni are essential to be input for theoretical models predicting the behavior under extreme neutron-to-proton ratios.
The spectroscopic measurements of $^{79}$Cu~\cite{Olivier2017,OlivierThesis} and $^{78}$Ni~\cite{Taniuchi2019,TaniuchiThesis} were conducted with a high-intensity cocktail beam at RIBF aiming at testing the persistence of the shell-closures and gaining an understanding of the structural description of extreme neutron-rich isotopes.

\subsection{\textsuperscript{79}Cu: a single-particle proton with the \textsuperscript{78}Ni core}
One of the most crucial conclusions to evaluate the persistence of the $Z=28$ shell gap in the region surrounding $^{78}$Ni was deduced by the detailed analysis of the $\gamma$-ray spectra of $^{79}$Cu~\cite{Olivier2017,OlivierThesis}.
$^{79}$Cu was produced through the direct knockout of a proton from $^{80}$Zn, one of the main species of the cocktail beam from the in-flight fission products of the \textsuperscript{238}U primary beam.

\begin{figure}
    \centering
    \includegraphics{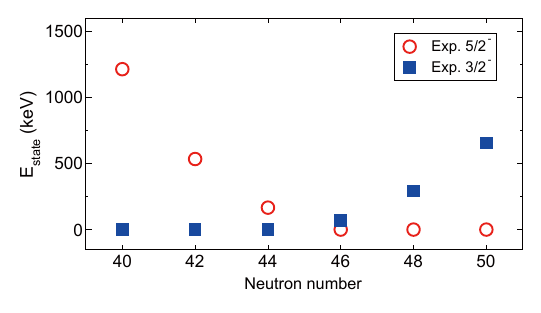}
    \caption{Energy of excited states presumed as single-particle states of protons for copper isotopic chain. The extension of the trend of the energy differences beyond the crossing point has been confirmed. Reprinted figure with permission from~\cite{Olivier2017}. Copyright 2024 by the American Physical Society.}
    \label{fig:cu_sp_states}
\end{figure}

The observed multiple excitation states around 3~MeV and the absence of significant feeding to excited states below $2.2~\textrm{MeV}$ upon the proton knockout reactions, suggest a strong spherical structure resistant to deformation, confirming the doubly magic nature of \textsuperscript{78}Ni.
One of the key aspects of the findings was the observation of systematic trends in the single-particle energies of the $5/2^-$ and $3/2^-$ states as the neutron number approached 50, as shown in Figure~\ref{fig:cu_sp_states}.
These trends indicated a continuation of the inversion of the $\pi p_{3/2}$ and $\pi f_{5/2}$ orbitals observed along the copper isotopes.
Such inversions are indicative of the effective single-particle energies with respect to neutrons filling the $pf$ shell.

The obtained results were in strong agreement with predictions made by Monte Carlo large-scale shell model calculations using the A3DA Hamiltonian~\cite{Tsunoda2014}, as discussed in Section~\ref{sec:MCSM}.
These calculations account for changes in the tensor forces as additional neutrons shift the energy levels of the proton orbitals.
The experimentally deduced exclusive cross sections agree with the theoretical predictions, which were obtained by multiplying the spectroscopic factors from the structure calculation with the single-particle cross sections provided by reaction calculations using the distorted-wave impulse approximation (DWIA)~\cite{Wakasa2017,Yoshida2024}.
The detailed level scheme elucidated in this study is vital for validating state-of-the-art theoretical models, particularly concerning the migration of single-particle states.

%

\subsection{\textsuperscript{78}Ni: the doubly-magic stronghold exhibiting shape coexistence}
An experimental investigation into the properties of $^{78}$Ni was conducted to confirm its character as a doubly magic nucleus~\cite{Taniuchi2019, TaniuchiThesis}.
In-beam $\gamma$-ray spectroscopy was conducted to measure the decay of excited $^{78}$Ni following proton knockout reactions. This approach enabled the first observation of its excited levels, providing an essential input for the state-of-the-art theoretical predictions.

\begin{figure}[p]
    \centering
    \includegraphics[width=\textwidth]{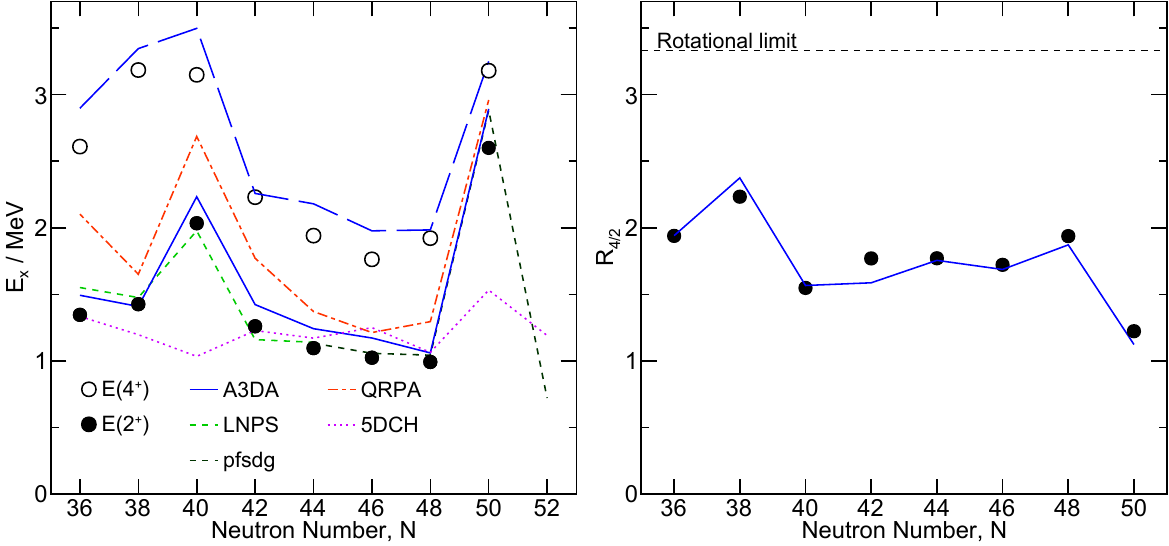}
    \caption{Systematic chart of excitation energies for the $2_1^+$ (filled circles) and $4_1^+$ (open circles) states, and their ratios $R_{4/2}$ along the nickel ($Z=28$) isotopic chain.
    This figure combines experimental data with theoretical calculations (lines), showcasing the evolution of these excited states as neutron number increases, and highlighting the effects of nuclear shell closures at $N=50$. The A3DA model calculated the $4^+$ trend well up to $N=50$. See the text for the explanation of each prediction.}
    \label{fig:E2-isotope}
    \vspace{5mm}\vfill
    \centering
    \includegraphics[width=\textwidth]{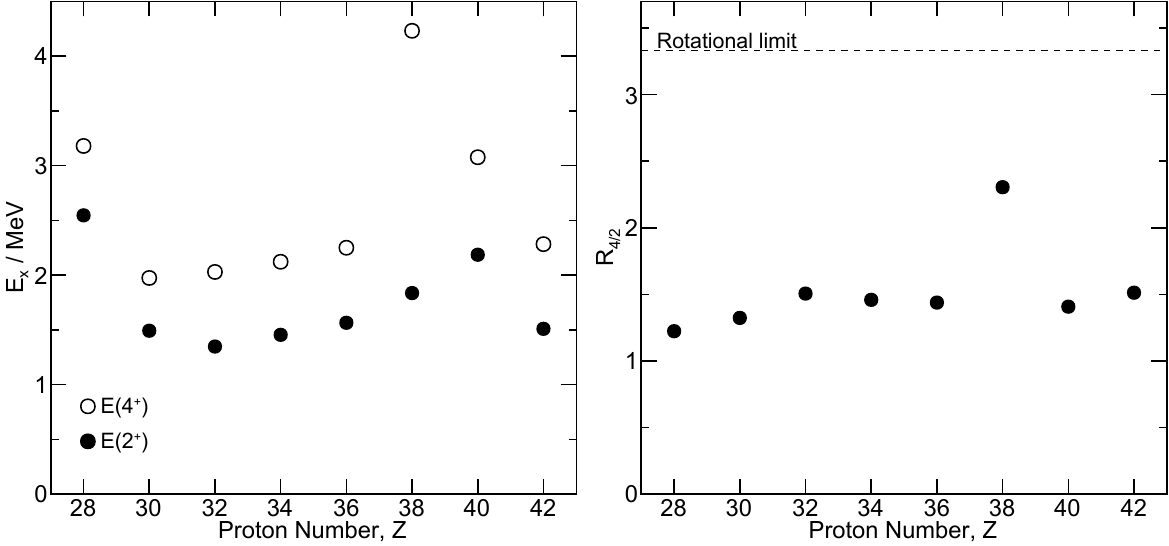}
    \caption{Excitation energies and their ratios $R_{4/2}$ along the isotonic ($N=50$) chain. The sudden increase of both $2^+$ and $4^+$ energies at $Z=28$ indicates the shell closure of \textsuperscript{78}Ni.
    The rotational limit at $R_{4/2}=10/3$ indicates when the system is excited as a rigid rotor.
    }
    \label{fig:E2-isotone}
\end{figure}

The energy of the first excited state was measured at 2.6~MeV, indicating the doubly magic nature of $^{78}$Ni.
This high excitation energy supports the robustness of the shell closure at this nucleus.
The systematic trends of the excitation energies along isotopic and isotonic chains, as shown in Figures~\ref{fig:E2-isotope} and \ref{fig:E2-isotone}, corroborate various theoretical models, which are explained in the following section.
Furthermore, another $2^+$ state at an energy of 2.9~MeV was identified in the two-proton knockout channel. This fact suggests the presence of shape coexistence.
This interpretation is supported by modern large-scale shell-model calculations, which include additional nucleon model spaces especially the neutron $sdg$ orbitals, potentially indicating a prolate shape coexisting with the spherical ground state.

The inclusive cross sections of one- and two-proton knockout reactions, $(p,2p)$ and $(p,3p)$, to produce \textsuperscript{78}Ni were found to be almost five times smaller than those of other isotopes in this region.
Theoretical cross sections were conducted using the DWIA calculation folding spectroscopic factors from several structure calculations.
The exclusive cross section populating the ground and $2_1^+$ states are reasonably explained by all the calculations.
The low inclusive cross section is considered due to the low neutron separation energy of \textsuperscript{78}Ni%
\footnote{%
$S_n=5.60(57)$~MeV with the latest atomic mass evaluation (AME2020)~\cite{Wang2021}. In the original paper of ref.~\cite{Taniuchi2019} used the older version of the evaluations, $S_n=5.16(78)$~MeV. The conclusion stays the same.
} while a significant portion of the populations is calculated to feed unbound states at higher energy.

\begin{figure}
    \centering
    \includegraphics[width=\textwidth]{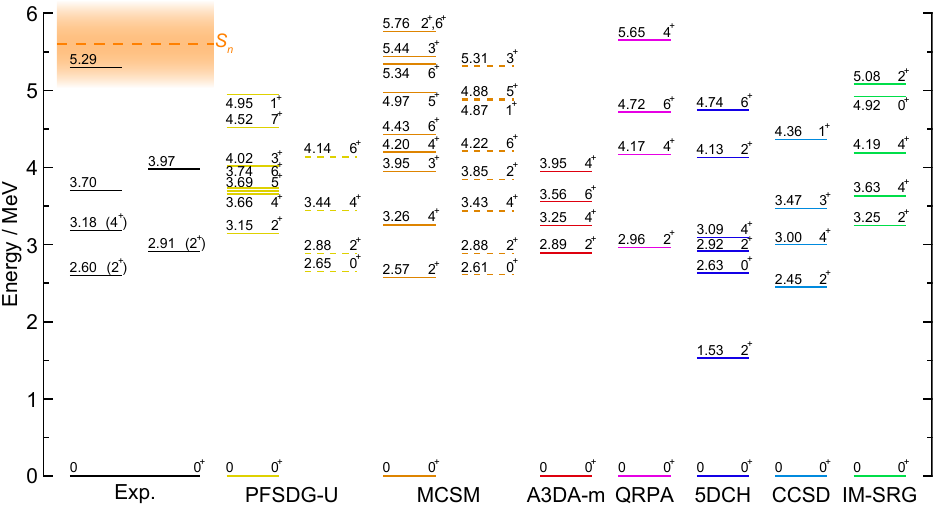}
    \caption{Energy levels of $^{78}$Ni as obtained experimentally~\cite{Taniuchi2019,TaniuchiThesis} compared with predictions from various theoretical models~\cite{Nowacki2016,Tsunoda2014,Hagen2016,Simonis2017,Delaroche2010,Peru2014}.}
    \label{fig:78NiLevels}
\end{figure}

Figure~\ref{fig:78NiLevels} illustrates the experimentally obtained energy levels alongside those predicted by various theoretical models, including those discussed in the following section.
Further investigations around these nuclei are essential for a comprehensive understanding of how the $N=40$ Island of Inversion extends towards and beyond the $N=50$ line, the driving force of the shell evolution, and the limit of the existence of neutron-rich isotopes.


\section{Theoretical works towards and beyond \textsuperscript{78}Ni}
Many theoretical investigations have been put in the effort to describe the nuclear structures around \textsuperscript{78}Ni. Here the studies employed in the $\gamma$-ray spectroscopy experiments are described and compared.
Firstly, two large-scale shell model calculations are discussed in Sections~\ref{sec:LSSM} and \ref{sec:MCSM}. Both predictions realized the necessity of including a large set of orbitals above $N=50$ to reproduce the shape coexistence in \textsuperscript{78}Ni.
Two first-principle frameworks, Coupled-Cluster (CC) and In-Medium Similarity Renormalization Group (IM-SRG) are described in Section~\ref{sec:abinitio}.
The beyond mean-field calculations, 5DCH and QRPA, are introduced in Section~\ref{sec:meanfield}. 
Comparison of these predictions and the observed energies are summarized in Figure~\ref{fig:78NiLevels}.



\subsection{Large scale shell-model calculations}\label{sec:LSSM}

Significant advancements in understanding the nuclear structure of \textsuperscript{78}Ni have been achieved through large-scale shell-model calculations, utilizing both the LNPS interaction~\cite{Lenzi2010, Santamaria2015} and the one with an extended model space known as PFSDG-U~\cite{Nowacki2016}.

\vspace{3mm}\noindent \textbf{LNPS}\quad
Based on a \textsuperscript{48}Ca core, the LNPS interaction includes the full $pf$ shell for protons and the orbitals $0f_{5/2}$, $1p_{3/2}$, $1p_{1/2}$, $0g_{9/2}$, and $1d_{5/2}$ for neutrons. This interaction is derived from a realistic nucleon-nucleon potential, with empirically corrected monopole parts to better align with experimental data, providing successful descriptions of the IoI at and beyond $N=40$.

\vspace{3mm}\noindent \textbf{PFSDG-U}\quad
The PFSDG-U interaction employs a larger model space that includes the full $pf$ shell for protons and full $sdg$ shell for neutrons. This extension requires more comprehensive interactions involving additional proton and neutron orbitals, crucial for reproducing the nuclear structure of \textsuperscript{78}Ni.

\begin{figure}[tb]
    \centering
    \includegraphics{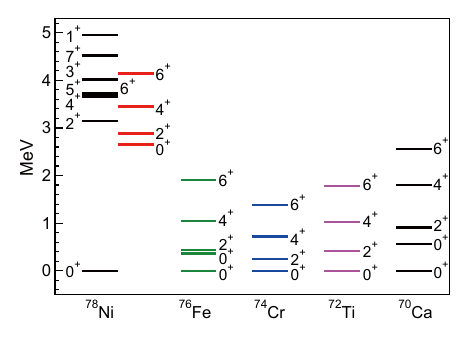}
    \caption{Energy levels predicted by the large-scale shell model calculation with the PFSDG-U interaction showing the evolution of the $N=50$ shell closure from \textsuperscript{78}Ni towards \textsuperscript{70}Ca.
    This figure highlights the breakdown of traditional shell closures and the onset of shape coexistence in \textsuperscript{78}Ni.
    Reprinted figure with permission from~\cite{Nowacki2016}. Copyright 2024 by the American Physical Society.}
    \label{fig:lssm_levels}
\end{figure}

\begin{figure}[tb]
    \centering
    \includegraphics[width=0.7\textwidth]{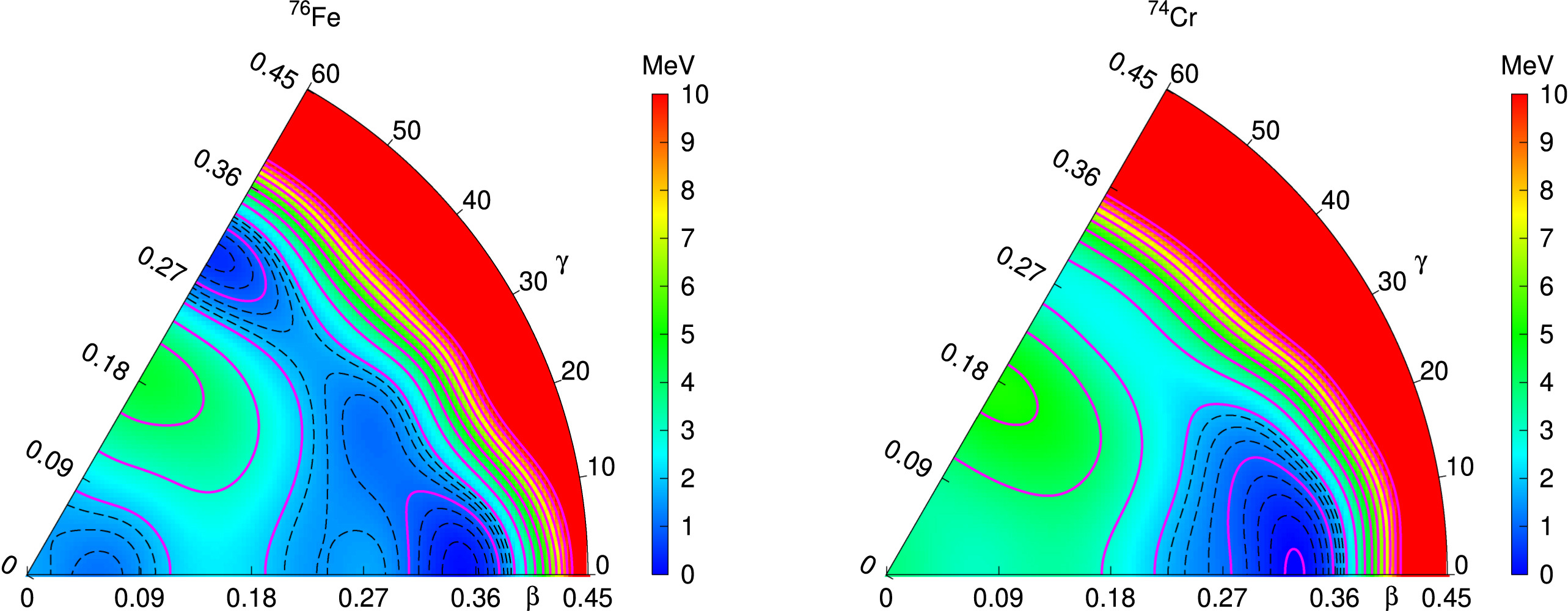}
    \caption{Potential Energy Surfaces (PES) of \textsuperscript{76}Fe (left) and \textsuperscript{74}Cr (right) predicted by LSSM with the PFSDG-U interaction. The figure is taken from ref.~\cite{Nowacki2021} under the CC BY-NC-ND license.}
    \label{fig:pes50}
\end{figure}

\vspace{3mm}
The shell-model calculations with the PFSDG-U interaction predicted the intensive competition between spherical and deformed configurations in \textsuperscript{78}Ni and its neighbors.
Following the number of protons decreases along the isotonic chain, deformed states become dominant rapidly.
Figure~\ref{fig:lssm_levels} illustrates the calculated levels by PFSDG-U interactions along the $N=50$ chain. For \textsuperscript{78}Ni, two distinct bands are depicted: one representing the traditional spherical shape associated with shell closure, and a second, higher-energy band suggesting a deformed configuration. This dual-band structure highlights the ongoing competition between the two nuclear shapes, with the spherical form maintaining dominance at the ground state.
Toward lighter isotones, such as \textsuperscript{76}Fe, the spherical and deformed states almost degenerate. 
In the more exotic isotones, \textsuperscript{74}Cr, \textsuperscript{72}Ti, and \textsuperscript{70}Ca, the calculations predict these nuclei to be predominantly deformed, indicating a clear breakdown of the traditional $N=50$ shell closure and a rapid transition to deformed ground states.
Figure~\ref{fig:pes50} depicts the Potential Energy Surfaces (PES) of these isotones at the transition to deformed shapes. While \textsuperscript{76}Fe has shape coexisting minima, \textsuperscript{74}Cr shows a well-developed prolate deformation.
This suggests extensions of the $N=40$ Island of Inversion, where increased collectivity and deformation are observed in chromium and iron isotopes, towards $N=50$~\cite{Santamaria2015}.

Recent studies, such as the detailed $\beta$-delayed $\gamma$-ray spectroscopy~\cite{Rocchini2023} and the precise mass measurements of the isomeric state of \textsuperscript{79}Zn$^m$~\cite{Nies2023}, compared with the latest calculations with PFSDG-U, further elucidate the possible shape coexistence in nuclei around \textsuperscript{78}Ni.
These findings affirm that \textsuperscript{78}Ni marks the onset of significant nuclear deformation, extending towards neutron-rich isotones, with the largest deformations expected at $N=56$~\cite{Nowacki2021}.

It is also worth mentioning the recent development of a diagonalization method using a discrete non-orthogonal basis (DNO-SM), as outlined by Dao and Nowacki~\cite{Dao2022}. This method uses an optimized set of constrained Hartree-Fock states as basis configurations, significantly reducing the number of basis states needed compared to conventional shell-model calculations. Benchmark results show that DNO-SM achieves an accuracy comparable to exact SM diagonalization, making it a powerful approach for capturing deformed nuclear shapes and shape coexistence across a wider mass range of isotopes.

\subsection{Monte Carlo Shell Model (MCSM) Calculations}\label{sec:MCSM}

The Monte Carlo Shell Model (MCSM) is an effective theoretical framework for exploring detailed structures of nuclei, particularly renowned for its ability to handle the diagonalization of very large matrices involving many valence nucleons. This capability allows MCSM to compute configurations across a broad range of single-particle orbitals, crucial for detailed predictions of complex nuclear structures and various excited states.

The A3DA-m interaction~\cite{Tsunoda2014} represents a significant advancement in MCSM to predict the neutron-rich nickel region.
It is defined with a core of \textsuperscript{40}Ca and includes valence orbitals within the $pf$ shell, and the $0g_{9/2}$ and $1d_{5/2}$ orbits for both protons and neutrons.
This interaction reproduces the nuclear structure and their shell evolution such as the shape coexistence around $N=40$ region, especially in \textsuperscript{68}Ni, while no coexisting shapes, but a rigid spherical structure of \textsuperscript{78}Ni is predicted, which is probably attributed by the limited model space for neutrons.

Further advancements in MCSM calculations for \textsuperscript{78}Ni involve an interaction, as discussed in the supplementary material of Taniuchi et al.~\cite{Taniuchi2019} with an extended model space of the full $pf$ and $sdg$ shells for both protons and neutrons.
The extension was found to be critical to reproduce the shape coexistence of \textsuperscript{78}Ni.
The new interaction offers predicting powers into the deformations and shell evolution towards further neutron-rich isotopes, which are essential for the reproductions of the r-process nucleosynthesis path and for accurately estimating the location of the neutron drip lines.
The findings underscore the necessity of incorporating broader model spaces to fully grasp the dynamics of nuclear structure and deformation, particularly for neutron-rich isotopes which require coverage across several major shells.

\subsection{First-principle calculations}\label{sec:abinitio}
The nuclear structure of neutron-rich isotopes near \textsuperscript{78}Ni has been intensively studied through first-principles approaches like Coupled-Cluster (CC)\cite{Hagen2016} and In-Medium Similarity Renormalization Group (IM-SRG)\cite{Simonis2017,Taniuchi2019}. Both frameworks predict nuclear structure based on the chiral interaction 1.8/2.0~(EM)~\cite{Hebeler2011}, which is well-regarded for accurately reproducing binding energies and energy spectra across a wide range of isotopes, from light to heavy masses.
While these frameworks were initially limited to spherical states when the first \textsuperscript{78}Ni spectroscopy data was published, they now have the capability to incorporate deformed harmonic oscillator bases in recent calculations. This section highlights recent advances in applying these methods to medium-heavy isotopes.

\begin{figure}
    \centering
    \includegraphics[scale=1]{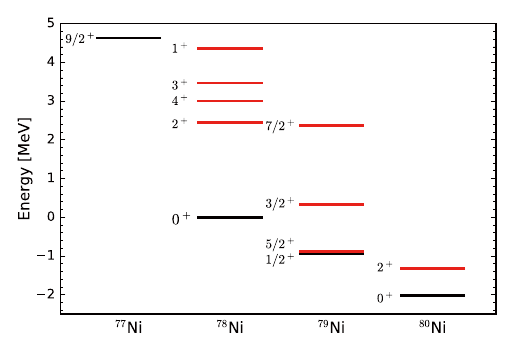}
    \caption{Predicted energy levels of \textsuperscript{77}Ni to \textsuperscript{80}Ni with the CCSD(T) calculations. Reprinted figure with permission from~\cite{Hagen2016}. Copyright 2024 by the American Physical Society.}
    \label{fig:cc_levels}
\end{figure}

\vspace{3mm}\noindent \textbf{Coupled Cluster}\quad
The Coupled Cluster with Singles and Doubles (CCSD) method predicts excited states by systematically including correlations up to 2p-2h excitations. To capture additional correlations, the CCSD(T) approach (CCSD with perturbative Triples) extends this by incorporating 3p-3h excitations perturbatively, reducing excitation energies by 1-2~MeV and significantly enhancing accuracy. This approach is particularly effective for predicting detailed properties near doubly closed-shell nuclei, such as \textsuperscript{78}Ni and \textsuperscript{48}Ca. Figure\ref{fig:cc_levels} illustrates energy levels predicted by CCSD(T) calculations applied to the neutron-rich nickel isotopes~\cite{Hagen2016}, successfully reproducing the first $2^+$ energy of \textsuperscript{78}Ni at 2.45~MeV and the nearby binding energy.

Recently, notable advances have been made in incorporating nuclear deformations into these frameworks, such as the development of a deformed coupled-cluster approach based on a deformed Hartree-Fock state~\cite{Novario2020, Hagen2022, Sun2024}. Although computationally demanding, these advancements pave the way for studying shape-coexisting nuclei as computational power continues to grow. Calculations of deformed states along the $N=50$ chain towards south from \textsuperscript{78}Ni reproduce low-lying rotational band states~\cite{Hu2024}, consistent with experimental observations and shell-model predictions.

\vspace{3mm}\noindent \textbf{IM-SRG}\quad
Unlike the Coupled Cluster method, the IM-SRG framework can handle larger model spaces, as demonstrated in its application to \textsuperscript{78}Ni using a \textsuperscript{60}Ca core~\cite{Taniuchi2019}. This expanded model space allows IM-SRG to more systematically predict isotopes around \textsuperscript{78}Ni. IM-SRG employs a unitary transformation to convert the many-body Hamiltonian into a diagonal or block-diagonal form, effectively decoupling the valence space from the remaining nuclear orbitals. This decoupling enables precise predictions of properties like two-neutron separation energies, charge radii, and excitation energies for a wide range of isotopes. For \textsuperscript{78}Ni, the IM-SRG(2) approximation was used, incorporating only 1p-1h and 2p-2h excitations. It is comparable to the relation between the CCSD against the CCSD(T) calculations, resulting in slightly higher predicted energies due to omitted correlations. Under these constraints, the predicted $2_1^+$ energy of 3.25~MeV is considered a reasonable estimate.

More recently, the valence-space density matrix renormalization group (VS-DMRG) has been applied to \textsuperscript{78}Ni and other medium-mass nuclei~\cite{Tichai2024}. This method provides a more accurate solution than the original IM-SRG calculations used in previous studies~\cite{Taniuchi2019}, which relied on approximate diagonalizations.

\vspace{3mm} %
Both CCSD(T) and IM-SRG employ chiral EFT interactions 1.8/2.0~(EM), yet they exhibit notable differences in their predictive capabilities. CCSD(T) enhances its precision in predicting excitation energies for \textsuperscript{78}Ni by incorporating 3p-3h excitations, while IM-SRG tends to yield higher energy predictions due to the absence of these higher-order correlations. Nevertheless, IM-SRG’s adaptability across a broader range of nuclear isotopes allows it to predict mass, charge radii, and excitation energies more comprehensively. Together, these methods provide complementary insights into nuclear structures for medium-mass isotopes, including \textsuperscript{78}Ni.

\subsection{Beyond Mean-Field Calculations}\label{sec:meanfield}
Beyond mean-field calculations, such as the Five-Dimensional Collective Hamiltonian (5DCH)~\cite{Delaroche2010} and Quasi-Particle Random Phase Approximation (QRPA)~\cite{Peru2014}, are applied to explore nuclear structures near \textsuperscript{78}Ni.

\vspace{3mm}\noindent \textbf{5DCH}\quad
The 5DCH approach is particularly effective for nuclei where deformation plays a dominant role in nuclear structure, especially for those away from closed shells.
By employing a triaxial deformed harmonic oscillator basis, 5DCH excels in calculating low-energy collective excitations.
However, for predominantly spherical nuclei like \textsuperscript{78}Ni, the approach is less effective. The excited states predicted by 5DCH for \textsuperscript{78}Ni are significantly underestimated, reflecting the strong shell closures and minimal deformation of its ground state.

\vspace{3mm}\noindent \textbf{QRPA}\quad
QRPA calculations primarily focus on vibrational excitations and do not account for rotational contributions, making them particularly suitable for describing nuclear structures in closed-shell isotopes.
While this method provides a reasonable description of the first $2^+$ states in nickel isotopes, it encounters difficulties in reproducing higher excited states, including the second $2^+$ state observed in \textsuperscript{78}Ni~\cite{Taniuchi2019}. 
 This limitation highlights the need for incorporating collective excitations associated with nuclear deformation to achieve a more comprehensive understanding of shape-coexisting nuclear structures, such as those in \textsuperscript{78}Ni.

\vspace{3mm}
The absence of the $2_2^+$ state in QRPA calculations, which reproduce only the first excited state under the assumption of spherical symmetry, indirectly supports the existence of shape coexistence in this region.



\section{Nuclear structures from $N=40$ towards $N=50$}
The origin of the possible shape coexistence emerged in \textsuperscript{78}Ni can be interpreted as the migration of the effective single-particle energies, that cause the collective excitations and the emergence of the deformed states.
A possible IoI at $N=50$ following the ones at $N=8$, 20, 28, and 40 is proposed~\cite{Nowacki2016}. Several experimental investigations on the possible quenching of the $N=50$ gap have been attempted from the ``North'' of the region along the $N=50$ isotones.

Besides that, the collapse of the $N=50$ shell is anticipated as a possible extension of the $N=40$ IoI from the ``Southwest'' of the nuclear chart, originally proposed by the LNPS interaction~\cite{Lenzi2010} and corroborated experimentally~\cite{Santamaria2015}. Here below, we show further detailed studies  to define the ``Northern-edge'' of the IoI approaching $N=50$ through several reaction channels around $Z=28$, proton inelastic scattering and spectroscopy of nickel and cobalt isotopes\footnote{%
Another review paper summarizing the experimental and theoretical works focusing on the $N=40$ IoI below $Z=28$ is published as another review paper in the same issue~\cite{CortesSua2024}.%
}.

\subsection{Inelastic Scattering of Neutron-Rich Ni and Zn Isotopes}
Inelastic scattering experiments using MINOS on neutron-rich \textsuperscript{72,74}Ni and \textsuperscript{76,80}Zn isotopes were conducted to examine their nuclear structure and reaction dynamics~\cite{Cortes2018, CortesThesis}.
These studies focused on measuring the angular integrated cross sections for direct inelastic scattering to the $2^+$ and $4^+$ states.
Hadronic scattering experiments, such as ($p,p^\prime$) complement the electromagnetic measurements by providing a probe for both proton and neutron excitations.
The experimental results were analyzed with the extended Jeukenne-Lejeune-Mahaux (JLM) folding model, integrating neutron and proton transition densities from QRPA calculations~\cite{Dupuis2016}.
The theoretical frameworks for interpreting inelastic nucleon scattering were developed to account for rearrangement corrections off spherical nuclei.
These corrections significantly influence the calculated cross sections, which vary depending on the incident energy and excitation multipolarities.

\begin{figure}
    \centering
    \includegraphics[scale=0.8]{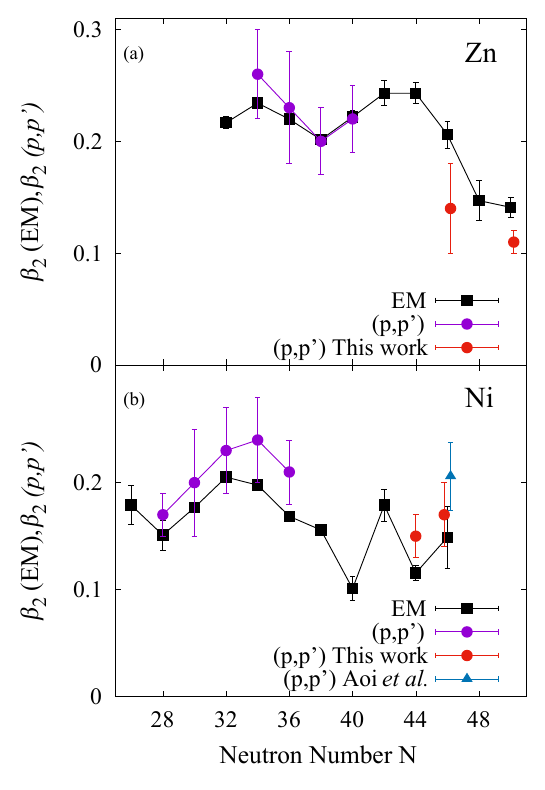}
    \caption{Systematic trend of the deformation parameters $\beta_2$ deduced from electromagnetic (EM) and hadronic ($p,p^\prime$) probes for (a) zinc and (b) nickel isotopes.
    Reprinted figure with permission from~\cite{Cortes2018}. Copyright 2024 by the American Physical Society.}
    \label{fig:syst_deform}
\end{figure}
Figure~\ref{fig:syst_deform} illustrates the systematic trend of the deformation parameters $\beta_2$ deduced from both electromagnetic (EM) probes of Coulomb excitations and hadronic ($p,p^\prime$) probes.
The comparison of these deformation parameters with $B(E2\uparrow)$ values enables assessing the degree of shell closures at the $Z=28$ and $N=50$ gaps for nickel and zinc isotopes.
This work revealed that the first $2^+$ states in nickel isotopes are predominantly influenced by neutron excitations, corroborating the $Z=28$ shell closure, while observations in zinc isotopes indicate significant proton excitations.
Additionally, these transition probabilities and the matrix elements are compared with both LSSM~\cite{Nowacki2016} (Section~\ref{sec:LSSM}) and QRPA~\cite{Peru2014} (Section~\ref{sec:meanfield}) calculations.
%

\subsection{Proton-hole states at highly excited states populated through \textsuperscript{77}Cu(p,2p)\textsuperscript{76}Ni}

The proton-hole states in \textsuperscript{76}Ni were extensively studied through the one-proton knockout reaction \textsuperscript{77}Cu($p,2p$)\textsuperscript{76}Ni~\cite{Elekes2019}, highlighting the detailed structure of the nickel isotopes as it approaches the $N=50$ neutron shell closure. This investigation took advantage of the high efficiency of the DALI2 detector array~\cite{Takeuchi2014} and the high luminosity of MINOS~\cite{Obertelli2014}, enabling the $\gamma\gamma$ and $\gamma\gamma\gamma$ coincidence analyses. With careful investigation, highly excited states around 4~MeV were identified.
The results were compared with the LSSM calculation employing the LNPS interaction~\cite{Lenzi2010} as discussed in Section~\ref{sec:LSSM}, and the DWIA calculation~\cite{Wakasa2017} for the exclusive cross sections. A good agreement between the theory and the observations was confirmed and reinforces the robustness of the shell closure at $Z=28$.
Although the analysis identified three new transitions, the detailed properties, such as spins and parities, are yet to be identified. Including the spectroscopy in neighboring odd-even isotopes, further experiments with a better resolving power detector are desired.

\subsection{Spectroscopy of \textsuperscript{69,71,73}Co via One-Proton Knockout Reactions}

\begin{figure}
    \centering
    \includegraphics[width=\textwidth]{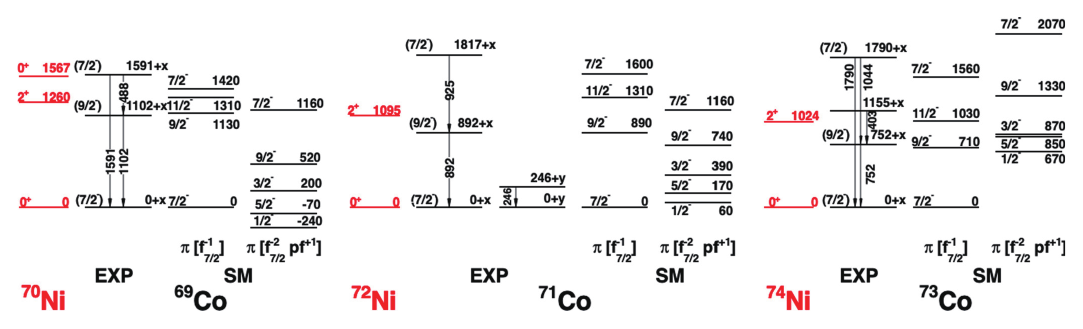}
    \caption{Measured excited states of \textsuperscript{69-73}Co in comparison with shell model calculations and the ones of \textsuperscript{70-74}Ni. Reprinted figure with permission from~\cite{Lokotko2020}. Copyright 2024 by the American Physical Society.}
    \label{fig:Lokotko}
\end{figure}

Spectroscopic studies of \textsuperscript{69,71,73}Co were conducted using the ($p,2p$) one-proton knockout reactions~\cite{Lokotko2020,LokotkoThesis}. These investigations utilized $\gamma\gamma$ coincidence analysis to construct the levels of the excited states. The identified excitation levels displayed in Figure~\ref{fig:Lokotko} are compared to the $2^+$ states of \textsuperscript{70,72,74}Ni. The $9/2_1^-$ and $11/2_1^-$ states in cobalt isotopes can be approximated as $\pi f_{7/2}^{-1} \otimes 2_1^+(\textrm{Ni})$. Additionally, large-scale shell model calculations, as described in Section~\ref{sec:LSSM}, employing the LNPS~\cite{Lenzi2010} and PFSDG-U~\cite{Nowacki2016} interactions, were used to calculate these excited states. The good agreement between experimental results and theoretical predictions suggests the coexistence of spherical and deformed shapes at low excitation energies within these cobalt isotopes, similarly as seen in \textsuperscript{78}Ni~\cite{Taniuchi2019}.

Moreover, the exclusive cross sections, $\sigma_\textrm{ex}$, were compared with calculated single-particle cross sections, $\sigma_\textrm{sp}$, using the DWIA~\cite{Wakasa2017} reaction model.
With potential unobserved feeding from higher excited states, the experimentally deduced exclusive cross sections are considered upper limits. The ratios of the cross sections, $\sigma_\textrm{ex}(g.s)/\sigma_\textrm{sp}$, are considered as spectroscopic factors and are found to be comparable with other measurements of the less neutron-rich isotopes \textsuperscript{61,63}Co. Notably, the observations of the second excited states of the spherical bands, exhibiting large cross sections contrasting with no direct population to the first excited states, corroborate the spin and parity assignments predicted by the shell-model calculations.


\section{Towards further understanding}
To advance our understanding of nuclear structure, particularly concerning shell evolution and shape competition, it is critical to extend investigations beyond the doubly magic stronghold of \textsuperscript{78}Ni. This involves exploring both neutron-rich isotopes such as \textsuperscript{79,80}Ni, \textsuperscript{77}Co, and \textsuperscript{76}Fe.

Advanced spectroscopic studies are necessary to overcome the limitations of earlier spectrometers like DALI2, which have modest energy resolution. The HiCARI experiment~\cite{Wimmer2021} at RIBF, utilizing high-purity Germanium (HPGe) detectors, was designed to significantly enhance resolution. This enhancement enables detailed lifetime measurements, providing further insights into excited states, including matrix elements and deformations.
Recently, experiments focusing on \textsuperscript{79}Cu and neutron-rich zinc isotopes~\cite{Taniuchi2022, Kaci2023} have been conducted, which are expected to provide further insights into the emerging shape coexistence in this region.

To observe shape coexistence more effectively and directly, developing new techniques, such as missing-mass spectroscopy, will be the key. These methods provide direct identification of $0_2^+$ states which often can not be concluded with $\gamma$-ray spectroscopy. The newly implemented STRASSE recoil-proton trackers~\cite{Liu2023} in inverse kinematics, combined with the HYPATIA scintillator array~\cite{Tanaka2024,hypatia_web} --- a high-efficiency and high-resolution $\gamma$-ray spectrometer --- will serve as an innovative and unique experimental setup for these studies, as highlighted in another review article in this issue~\cite{Tanaka2024}. These instruments are specifically suited to facilitate the study of ``hidden'' $0_2^+$ states and the detection of the weak transitions from the excited states in the collective bands around \textsuperscript{78}Ni and other short-lived neutron-rich isotopes.

While extensive spectroscopic studies should be pursued, mass measurements of copper and nickel isotopic chains beyond $N=50$ equally take an important role in deducing the gaps of the $N=50$ and $Z=28$ shells and for the precise determination of any existing isomeric states.
Furthermore, detailed investigations of the charge and matter radii of these nuclei will be important to be an input for theoretical calculations.

In addition to these experimental setups, studies of reaction kinematics, particularly using the two-proton knockout ($p,3p$) channel, will be the key to exploring potential deformed states in \textsuperscript{78}Ni and its neighbors. The development of microscopic reaction theories capable of handling four or more participants, extending beyond the DWIA framework that successfully reproduces ($p,2p$) reactions, is necessary.
Moreover, other reaction channels, such as inelastic scattering reactions like Coulomb excitation and ($p,p^\prime$) reactions, and transfer reactions at low-energy beams --- including those at ISOL facilities, re-accelerated, and slowed-down beams at secondary beam facilities --- will deepen our understanding of single-particle and collective states.

To complement these experimental efforts, advancements in nuclear structure calculations are also essential. Developing effective shell-model calculations that can handle larger model spaces is fundamental for accurately predicting and interpreting the complex structures of neutron-rich isotopes. Advances in first-principle calculations that incorporate different shape states and many-particle many-hole excitations, along with enhancements in computational power, would significantly enhance our understanding of nuclear structures across a broad range of the nuclear chart. Furthermore, integrating beyond mean-field approaches to better handle spherical and deformed excitations simultaneously will improve predictions and interpretations of experimental results.

\section{Summary}
Intensive and collaborative studies near \textsuperscript{78}Ni, employing state-of-the-art experimental devices and cutting-edge theoretical frameworks, have significantly advanced our understanding of nuclear structure in this region.
Although \textsuperscript{78}Ni was originally considered a doubly magic isotope, recent studies have revealed it as a focal point where shell closure competes with deformation.
Over the past decade, significant milestones have been achieved, including the confirmation of the doubly magic nature of \textsuperscript{78}Ni and its neighboring isotopes, as well as the identification of shape coexistence through intensive spectroscopy and inelastic scattering experiments.
Continued research in this region will not only deepen our understanding of nuclear structure but also lead us to investigate the driving forces behind shell evolution and shape coexistence in neutron-rich and weakly bound isotopes.
Moreover, these efforts will contribute to our understanding of the rapid nucleosynthesis processes where extremely neutron-rich isotopes take major roles.

\section*{Acknowledgment}
I thank F.~Nowacki, S.~Franchoo, Y.~Tsunoda, and Z.~Elekes for the discussions and their contributions to the dedicated sessions during the workshop held in York, and S.~Chen, M.~L.~Cort\'es, G.~Hagen, T.~Miyagi, F.~Nowacki, and A.~Obertelli for their valuable comments in preparing this review paper.
I acknowledge support from the UK STFC and the Royal Society.
On be half of the conference organizers, I appreciate the financial supports for the event from JSPS London, IoP, and EPJ.


\bibliographystyle{ptephy}
\bibliography{manualref.bib}


\let\doi\relax



\end{document}